\documentstyle[preprint,aps]{revtex}

\begin{document}
\draft

\title{The three-baryon $\Lambda$NN potential}

\author{A. A. \ Usmani}
\address{Interdisciplinary Laboratory, SISSA, via Beirut 2-4, 34014-Trieste,
Italy
}

\date{March 4, 1994}
\maketitle
\begin{abstract}
Using a three-baryon $\Lambda NN$ potential
having a spin-independent dispersive and a two-pion exchange components,
realistic variational Monte Carlo calculations have been performed for the
 $\Lambda$-seperation energy for $^{5}_{\Lambda}$He.
A relationship between strengths associated with both pieces of $\Lambda NN$
force
 giving exact experimental $\Lambda$-seperation energy has been obtained
which then combined with our previous study on $^{17}_{\,\,\Lambda}$O
hypernucleus determines a unique and acceptable set of strength parameters for
$\Lambda NN$ potential. The $\Lambda NN$ force is found important for
the core polarization.

\end{abstract}
\pacs{\ \ \ PACS numbers: 21.80.+a,21.10.Dr,13.75.Ev,27.20.+n}

\newpage
The Cluster Monte Carlo(CMC) technique first used for
nuclei\cite{steve} is recently generalised to $^{17}_{\,\,\Lambda}$O
hypernucleus for a realistic microscopic calculations \cite{anisul}.
  The same has been further extended to $^{5}_{\Lambda}$He in this work
to study the A-dependence of the expectation value of
 $\Lambda NN$ potential which is of great importance.

In case of $^{17}_{\,\,\Lambda}$O,
we find that the operatorial $NN,\ NNN,\ \Lambda N$ and $\Lambda NN$
correlations completely change the previously determined $\Lambda NN$
potential using central correlations\cite{bodmer} and that a reasonable
$V_{\Lambda NN}$ results in a $\Lambda$-seperation energy$(B_{\Lambda})$
consistent with the "empirically" determined value i.e. -13 $\pm 0.4$ MeV.
Also, $\Lambda NN$ force generates a corresponding correlation which has
been found playing a significant role in changing the density profile and
energy of the core nucleus. This provokes a deep interest in determining
the correct three-baryon $\Lambda NN$potential.

The $\Lambda NN$ potential is written as a combination of a dispersive
central force\cite{bodmer} ($V_{\Lambda NN}^{D}$) and a two-pion exchange
force\cite{bhaduri} ($V_{\Lambda NN}^{2\pi}$)
\begin{equation}
V_{\Lambda NN}= V_{\Lambda NN}^{D}+V_{\Lambda NN}^{2\pi}.
\end{equation}
	We do not invoke a spin dependence in the $V_{\Lambda NN}^{D}$  which
is unimportant for spin-zero core hypernuclei. Thus, we use a phenomenological
form
\begin{equation}
V_{NN\Lambda }^{D}=W_{0}T_{\pi}^{2}(r_{1\Lambda})T_{\pi}^{2}(r_{2\Lambda}).
\end{equation}
Indeed, we find a negligibly
small value for spin-spin $\Lambda N$ potential in $^{17}_{\,\,\Lambda}$O
\cite{anisul} and also herein for $_{\Lambda}^{5}$He.
 $T_{\pi}$(r) is the OPE tensor potential
\begin{equation}
T_{\pi}=\left({1+{3\over x }+{3\over x^{2}}}\right) {e^{-x}\over x}
\left({1-e^{-cr^{2}}}\right)^{2}\,
\end{equation}
with x=$\mu$r, $\mu$=0.7 fm$^{-1}$ is the pion mass,
and the cut-off parameter c=2.0 fm$^{-2}$.
The $V_{\Lambda NN}^{2\pi}$, consists of two parts, arising from s- and p- wave
$\pi -\Lambda$ potentials\cite{bhaduri}. The former is strongly suppressed
 and that almost all contribution is coming from the latter one. We, thus,
do not consider the s-wave channel and write the $V_{\Lambda NN}^{2\pi}$ as

\begin{equation}
V_{\Lambda NN}^{2\pi}=-\left({C_{p}\over 6}\right)
(\mbox{\boldmath$\tau$}_{1} \cdot
\mbox{\boldmath$\tau$}_{2}) \left\{{\mbox{\boldmath$\sigma$}_{1}
\cdot\mbox{\boldmath$\sigma$}_{\Lambda}
Y_{\pi}(r_{1\Lambda})+S_{1\Lambda}T_{\pi}(r_{1\Lambda}),
\mbox{\boldmath$\sigma$}_{2}.\mbox{\boldmath$\sigma$}_{\Lambda}
Y_{\pi}(r_{2\Lambda})+S_{2\Lambda}T_{\pi}(r_{2\Lambda})}\right\},
\end{equation}
The $S_{i\Lambda}$ is the tensor operator and
\begin{equation}
Y_{\pi}(r)={e^{-\mu r}\over \mu r} \left({1-e^{-cr^{2}}}\right).
\end{equation}
The experimental $B_{\Lambda}$ value has been obtained theoretically
by the relation
\begin{equation}
B_{\Lambda}={{<\Psi_{N}|H_{N}|\Psi_{N}>}
\over {<\Psi_{N}|\Psi_{N}>}}- {{<\Psi |H|\Psi >} \over
 {<\Psi |\Psi >}},
\end{equation}
where, $\Psi$ is the full wavefunction of the hypernucleus, $\Psi_{N}$ is the
ground state wavefunction of A-1 nucleons. The H and $H_{N}$  are the
hamiltonians for the hypernucleus and its core,
\begin{equation}
H=H_{N} \ -{\hbar^{2} \over {2m_{\Lambda}}}
\mbox{\boldmath$\nabla$}_{\Lambda}^{2} \ +\sum_{i=1}^{A-1}
v_{i\Lambda} \ +\sum_{i<j}^{A-1}
V_{\Lambda ij},
\end{equation}
\begin{equation}
H_{N}=-\displaystyle\sum_{i=1}^{A-1} {{\hbar^{2}} \over {2m}}
\mbox{\boldmath$\nabla$}_{i}^{2} \ +\sum_{i<j}^{A-1}
v_{ij} \ +\sum_{i<j<k}^{A-1}V_{ijk}.
\end{equation}
where $v_{ij}$ and $V_{ijk}$ refer to  NN Argonne potential\cite{wiringa1}
and NNN Urbana model VII potential\cite{schiavilla}.
The two-baryon $V_{\Lambda N}$ force is written as a sum of purely central,
space exchange and spin-spin pieces as
\begin{equation}
v_{\Lambda N}(r)=v_{0}(r)(1-\epsilon+\epsilon P_{x})+{1\over 4}
v_{\sigma}T_{\pi}^{2}(r)\mbox{\boldmath$\sigma$}_{\Lambda}\cdot
\mbox{\boldmath$\sigma$}_{N},
\end{equation}
\begin{equation}
v_{0}(r)=v_{c}(r)-\bar{v} T_{\pi}^{2}(r),
\end{equation}
where $P_{x}$ is the space exchange operator
and $\epsilon$ is the corresponding exchange parameter.  The $\bar{v}\equiv
(v_{s}+3v_{t})/4$
and $v_{\sigma}\equiv v_{s}-v_{t}$ are respectively the spin-average and
spin-dependent
strengths, where $v_{s}$ and $v_{t}$ denote the singlet and triplet state
depths,
respectively.
Finally, $v_{c}$(r) is a short range Saxon-Wood repulsive potential.
The various parameters are
$v_{s}$=6.33\,MeV,\,\, $v_{t}$=6.1\,MeV,\,\, $\epsilon$ =0.3,\,\,
$W_{c}$=2137\,MeV,\,\, R=0.5\,fm,\,\ and  a=0.2\,fm\,.
The variational wavefunction for $_{\Lambda}^{5}$He under the independent
triplet approximation is written as

\begin{equation}
|\Psi >  =
\left[\prod_{i=1}^{A-1} {(1+U_{i\Lambda})}\right] \ \left[S\prod_{i<j}^{A-1}
{(1+U_{ij})}\right] \ \left[1+\sum_{i<j}U_{ij\Lambda}+\sum_{i<j<k}U_{ijk}
\right] |\Psi_{J}>,
\end{equation}
\begin{equation}
|\Psi_{J}>=\prod_{i=1}^{A-1} f_{c}^{\Lambda}(r_{i\Lambda})\prod_{i<j}^{A-1}
 f_{c}(r_{ij}){\bf A} |\Phi^{A-1}>.
\end{equation}
The product ${\bf A}|\Phi^{A-1}>$ is a Slater determinant, where {\bf A}
is the antisymmetrization operator running over nucleons
\begin{equation}
{\bf A}|\Phi^{A-1}> = {\bf A} \left[{(n\uparrow)_{1}
(n\downarrow)_{2} (p\uparrow)_{3} (p\downarrow)_{4} (\Lambda\uparrow )_{5}}
\right].
\end{equation}
The three-baryon correlations
\begin{equation}
       U_{ijB} = \varepsilon_{B}V_{ijB},
\end{equation}
have the same structure as the three-baryon potentials though it differ
through the range c of the cut off functions $T_{\pi}$(r) and $Y_{\pi}$(r).
The subscript B stands for N and $\Lambda$ both.
 For $U_{ij}$ and $U_{i\Lambda}$ we make the choice

\begin{equation}
U_{ij}=\sum_{p=2}^{n} \beta_{p} u_{p}(r_{ij})O_{ij}^{p},
\end{equation}
\begin{equation}
 U_{i\Lambda}=\alpha_{\sigma}u_{\sigma}^{\Lambda}(r_{i\Lambda})
 \mbox{\boldmath$\sigma$}_{\Lambda}\cdot \mbox{\boldmath$\sigma$}_{i},
\end{equation}
with
\begin{equation}
u_{\sigma}^{\Lambda}=
{{f_{t}^{\Lambda}-f_{s}^{\Lambda}}\over f_{c}^{\Lambda}}.
\end{equation}
The $f_{c}^{\Lambda}$ is the spin averaged correlation function
\begin{equation}
f_{c}^{\Lambda}={{f_{s}^{\Lambda}+3f_{t}^{\Lambda}}\over 4}
\end{equation}
and $f_{s}^{\Lambda}$ and $f_{t}^{\Lambda}$ are the solutions of Schrodinger
 equations  with quenched $\Lambda$N potentials in singlet and triplet states
 respectively
\begin{equation}
\left[{-{\hbar^{2}\over 2\mu_{\Lambda N}}\mbox{\boldmath$\nabla$}^{2}
 +{\tilde{v}}_{s(t)}(r_{\Lambda N})
 +\theta_{\Lambda N}}\right] f^{\Lambda}_{s(t)} =0.
\end{equation}
The quenched potential ${\tilde{v}}_{s(t)}$ is quenched in the two-pion
 parts of the central and spin channels
\begin{equation}
 \tilde{v}_{s}(r)=\alpha_{\sigma}v_{c}(r)-
(\alpha_{2\pi}\bar{v}+\frac{3}{4}\alpha_{\sigma}v_{\sigma})T_{\pi}^{2}(r),
\end{equation}
\begin{equation}
 \tilde{v}_{t}(r)=\alpha_{\sigma}v_{c}(r)-(\alpha_{2\pi}\bar{v}-
\frac{1}{4}\alpha_{\sigma}v_{\sigma}T_{\pi}^{2}(r),
\end{equation}
and
\begin{eqnarray}
\theta_{\Lambda N}&=&{\hbar^{2}\over 2\mu_{\Lambda N}}
\left({\kappa^{2}_{\Lambda N}-
{2\kappa_{\Lambda N}(\nu_{\Lambda N}-1)
\over r}+{\nu_{\Lambda N}(\nu_{\Lambda N}-1)
\over r^{2}}}\right)\left({1-e^{-r^{2}
\over C_{\Lambda N}^{2}}}\right)
\nonumber \\
& &+{\gamma_{\Lambda N}\over \left({1+e^{r-R_{\Lambda N}
\over a_{\Lambda N}}}\right)}.
\end{eqnarray}
The $\theta_{\Lambda N}$ is such that the $\Lambda$-nucleon correlations in
eq(11)
 has the asymptotic behaviour required by the A-1 body Schrodinger equation

\begin{equation}
     f_{\Lambda N}^{c} \ \sim \ r^{1\over A-1} e^{-\kappa_{\Lambda N}r}.
\end{equation}
Also in $U_{ij}$,
a central three-body correlation is folded  as
\begin{equation}
U_{ij} \ \rightarrow \ \prod_{k\neq i,j}f_{ijk}U_{ij}.
\end{equation}
We refer to ref.\cite{wiringa2} for details of $f_{ijk}, U_{NN}$ and $U_{NNN}$
correlation functions.

	First of all, we made variational searches for $^{4}$He
of ref.\cite{wiringa2} and find that every parameter turns out to be the same
except the $\varepsilon_{N}$ which is -.005 instead of -.004. Then we made
variational parameters searches for $_{\Lambda}^{5}$He for many combinations
of $W_{0}$ and $C_{p}$ including $W_{0}=C_{p}=0$ (which means $U_{\Lambda NN}$
and $V_{\Lambda NN}$ have been set off). we find that $\alpha_{\sigma}$ is
unity
,however, unlike $_{\,\,\Lambda}^{17}$O, $\alpha_{2\pi}$ is found
sensitive to $C_{p}$ obeying
\begin{equation}
\alpha_{2\pi}=0.92-0.1\,C_{p}\,,\,\,\,\ 0\leq C_{p} \leq 0.7 \ $ MeV$\,.
\end{equation}
We also note an increase in $\alpha_{2\pi}$ from its minimum value at
$C_{p}=0.7$
\begin{equation}
\alpha_{2\pi}=0.85+0.2\,C_{p}\,,\,\,\,\ 0.7< C_{p}\leq 1.0 \ $MeV$\,.
\end{equation}
This is probably because $U_{\Lambda NN}$ and the repulsive $f_{c}^{\Lambda}$
both go together to suppress the nucleon density in the interior region unless
the most stable position is achieved and then $f_{c}^{\Lambda}$ gets increased
to cancel  the effects of $U_{\Lambda NN}$ with further enhancement of $C_{p}$.
In case of $_{\,\,\Lambda}^{17}$O this sentivity was weak. The parameter
$\varepsilon_{\Lambda}$ is found to be -.0014 and the cut off for
$U_{\Lambda NN}$ is 1.6 fm$^{-2}$.  The variational parameters used in
eq.(22) are
$\kappa_{\Lambda N}$=0.02\,fm,\,\, $\nu_{\Lambda N}$=0.7\,\,\,
$C_{\Lambda N}$ =2.3\,$fm^{-2}$,\,
$a_{\Lambda N}$=1.4\,fm,\,\, $r_{\Lambda N}$=0.7\,fm,\,\
and  $\gamma_{\Lambda N}$ is eigen value parameter determined by matching
logarithmic derivatives at a suitable r.
The $^{4}$He energy in our calculation corresponds to the first six operators
of Argonne $v_{14}$ potential and is 31.1$\pm 0.1$ MeV. The CMC technique up
to the four-body clusters
is complete for the core with four nucleons ,however, we neglect a sole
five-body cluster involving a $\Lambda$.

We observe that $V_{\Lambda NN}^{2\pi}$ is substantial and negative and that
it is due to the presence of non central correlations in the wave function
as its values are -2.65 $\pm 0.05$ and 0.19 $\pm$ 0.03 with and without non
central correlations, respectively. We also note that $S_{12}$ operator in
$U_{12}$ changes the $<V_{\Lambda 12}^{2\pi}>$. These results are along the
 line of $^{17}_{\,\,\Lambda}$O.

	We find a strong dependence of $B_{\Lambda}$ values on the
$\Lambda NN$ potential which has been fit in a functional form depending
upon the strengths $W_{0}$ and $C_{p}$ as

\begin{equation}
B_{\Lambda}(_{\Lambda}^{A}Z)=a+bC_{p}+cC_{p}^{2}+dW_{0}.
\end{equation}
The dependence on $W_{0}$ is linear, however, the quadratic
dependence on $C_{p}$ is attributed to the $U_{\Lambda NN}$ correlation.
The constants a, b, c and d depend upon the hypernucleus $_{\Lambda}^{A}Z$. The
constant a refers to the theoretically calculated $B_{\Lambda}$ value when
$V_{\Lambda NN}^{2\pi}$ force and its corresponding correlation is set off.
We variationally determine  $B_{\Lambda}$  for many combinations
of $W_{0}$ and $C_{p}$. For example, for $W_{0}=C_{p}=0
$, $B_{\Lambda}$($_{\Lambda}^{5}$He) is $4.8\pm 0.1$ and for
$W_{0}=.016, C_{p}=0.7, B_{\Lambda}$ is
$2.9\pm 0.1$. we then perform correlated difference Monte Carlo calculations
using the $W_{0}=.016, C_{p}=0.7$ random walk for other sets of
$W_{0}$ and $C_{p}$. We fit all these $B_{\Lambda}$ and
$\delta B_{\Lambda}$(given in table I) values using above functional form and
get a=4.8 MeV, b=2.45, c=2.5 MeV$^{-1}$ and d=-291.
Using these constants, one may then obtain a
relationship between $W_{0}$ and $C_{p}$ for the experimental $B_{\Lambda}$
($_{\Lambda}^{5}$He) i.e. 3.12$\pm 0.02$ MeV\cite{juric}.
A similar relationship between $W_{0}$ and $C_{p}$  has already been obtained
for $^{17}_{\,\,\Lambda}$O hypernucleus for its empirically etermined
$B_{\Lambda}$ value.
We then plot $W_{0}$ and $C_{p}$ for both the hypernuclei and get
a unique combination of $W_{0}$ and $C_{p}$ where the two plots intersect
 each other (fig. 1). This determines the $\Lambda NN$ potential with strengths
$W_{0}=0.015$ MeV  and $C_{p}=0.67$ MeV. Using these strengths, we reperform
the calculations and get $B_{\Lambda}$($_{\Lambda}^{5}$He)=3.1$\pm$0.1 MeV.
The core energy is -23.4 $\pm$ 0.2 MeV in this case. However,
 it is -21.0 $\pm$ 0.2 MeV with no $U_{\Lambda NN}$ and $V_{\Lambda NN}$ in
the calculation. Thus a distortion of 2.4 MeV is found due to ${\Lambda NN}$
 potential.  The results are given in Table II and in Table III for
the newly determined potential and for $W_{0}=C_{p}=0$, respectively.
 Accidently, we chose $W_{0}=.015$
and $C_{p}=0.7$ for the $_{\,\,\Lambda}^{17}$O in our previous
study\cite{anisul}
 which are very close to the newly determined strengths and had got the
$B_{\Lambda}(_{\Lambda}^{17}$O)=13.5$\pm 1.8$
MeV slightly higher than the empirical $B_{\Lambda}$. Certainly, a
decrease in $C_{p}$ from 0.7 to 0.67 will reduce the negative contribution
of $V_{\Lambda NN}^{2\pi}$ and will bring the results very close to the
emperical value. Thus, the conclusions drawn over there are correct for the
new $V_{\Lambda NN}$ potential too.

It is obvious from fig.2 (where point nucleon and $\Lambda$ densities are
shown)
that $\Lambda$ is strongly localised in the interior region of the core and
that it pushes away the nucleons towards lower density regions both periphery
and centre of the nucleus, presumably due to repulsive $f_{\Lambda N}$
correlation. The
$U_{\Lambda NN}$ reduces the $\Lambda$ density in the interior region and thus
as a result nucleons are less pushed away.  In both cases of nucleon and
$\Lambda$ densities, $V_{\Lambda NN}$ and its corresponding correlation
$U_{\Lambda NN}$ are found to play a significant role.

The addition of the space-exchange correlation in $U_{i\Lambda}$ (eq.16)
is very unlikely to change the situation as it is weak and that the cut of
the curves in fig.1 is sharp. We are in a way to include this to see its
effect specially on the $_{\,\,\Lambda}^{17}$O hypernucleus where many p-wave
nucleons are present. Also, the use of spin dependent $V_{\Lambda NN}^{D}$
for spin non-zero core hypernuclei will be interseting. Such a study for
$_{\Lambda}^{6}$Li, $_{\Lambda}^{6}$He and $_{\,\,\Lambda}^{16}$O is in
progress. We will also try to include the s-wave part of the $V_{\Lambda NN}$.
Moreover, we believe that $U_{\Lambda NN}$ should be
 responsible for the $\Lambda$ spin-orbit splittings in various excited states
of hypernuclei as there is no two-body $\Lambda N$ tensor correlation. Using
the newly determined $C_{P}$ we will be able to calculate it for the
$_{\,\,\Lambda}^{16}$O for the $\Lambda$ being in p-shell with J=3/2 and 1/2.
The experimental $B_{\Lambda}(_{\,\,\Lambda}^{16}$O) for $\Lambda$ being in
p-shell is 2.5 $\pm$ 0.5 MeV\cite{millener}. Thus the splitting is less than
1 MeV. In other words, this study will provide a check on $C_{p}$.

The author acknowledges S. C. Pieper for many useful comments, suggestions
and discussions. He is also acknowledging Professor A. Zichichi, ICSC-World
Laboratory, Lausanne, Switzerlad for financial support,
 Professor S. Fantoni, Interdisciplinary Laboratory, SISSA, Trieste, for
 encouragements and R B Wiringa for providing clarifications on his
$^{4}$He results.  Calculations have been performed on the computers
available at SISSA.

\begin{figure}
\caption{$C_{p}$ vs $W_{0}$.}
\end{figure}

\begin{figure}
\caption{One-body densities for the nucleons and $\Lambda$ in
$^{5}_{\Lambda}$He,
and for $^{4}$He.}
\end{figure}

\newpage
\begin{table}[t]
\caption{Differences $B_{\Lambda}(C_p,W_0)$-$B_{\Lambda}(C_p = 0.7, W_0 =
0.016)$.
}
\begin{tabular}{ldc}
\hline   \\
$C_p$				&	$W_0$								&	$\delta B_{\Lambda}$ MeV \\
\\
\hline \\
0.7					       &0.02					             				            &$-0.98 \pm 0.03$
   \\
0.7					       &0.018					            			            &$-0.49 \pm 0.01$
  \\
0.7					       &0.014					             			            &$+0.5 \pm 0.02$
  \\
0.8					       &0.02					            			            &$-0.46 \pm 0.07$
 \\
0.8					       &0.016					            		            &$+0.71 \pm 0.05$
 \\
0.8					       &0.014					             		            &$+1.31 \pm 0.05$
  \\
1.0					       &0.022					            		            &$+0.21 \pm 0.1$
\\
1.0					       &0.018					             		            &$+1.54 \pm 0.1$
 \\
1.0					       &0.016					             		            &$+2.22 \pm 0.1$
 \\
\\
\hline
\end{tabular}
\end{table}

\newpage
\begin{table}[t]
\caption{Variational Energies for $C_p$ = 0.67 MeV, $W_0$ = 0.015 MeV.  All
values are in
MeV.  Numbers in parentheses are statistical errors in the last digit.}
\begin{tabular}{lcddddc}
\hline   \\
Clusters  &        & one-body   & two-body   &  three-body           &
four-body			& Total \\
\\
\hline \\
$\Lambda$ Kinetic Energy$^{\rm a}$
&           &11.6(1)	&0.02(1)	&1.08(4)	&0.03(2)	&12.76(8) \\
$\Lambda$N Potential	&$v_{0}(r)(1-\epsilon)$
											&&-17.10(9)	&-0.63(2)	&-0.03(3)	&-17.76(10) \\
&$v_{0}(r)\epsilon P_{x}$	&&-7.09(4)	&0.50(1)	&0.09(1)	&-6.50(4) \\
&$\frac{1}{4} v_{\sigma} T^{2}_{\pi} {\mbox{\boldmath$\sigma$}}_{\Lambda} \cdot
{\mbox{\boldmath$\sigma$}}_{N}$		&
&0.09	&-0.03	&0.01	&0.07 \\
$\Lambda$NN Potential	&$V^{D}_{\Lambda NN}$
						&&&		3.23(3)	&0.11(1)	&3.34(3) \\
& $V^{2\pi}_{\Lambda NN}$	&&& -2.30(3)	&-0.36(2)	&-2.66(4) \\
$\Lambda$ Energy	&&11.6(1)	&-24.1(2)	&1.9(1)	&-.15(3)	&-10.7(1) \\
Nuclear Kinetic	&&81.1(3)	&47.0(1)	&7.6(1)	&-.3(1)	&135.4(4) \\
$NN$ Potential	&$v_{ij}$  &&-144.9(3)	&-3.5(1)	&1.2(1)	&-147.2(3) \\
$NNN$ Potential	&$V_{ijk}$  &&&-11.3(1)	&-.28(3)	&-11.6(1) \\
Nuclear Energy &&81.1(3)	&-97.9(2)	&-7.3(1)	&.6(1)	&-23.5(1) \\
Total Energy	&&92.7(3)	&-122.0(3)	&-5.4(1)	&.4(3)	&-34.2(1) \\
\\
\hline
\end{tabular}
\tablenotetext[1]{Includes nucleon kinetic energy from $\Lambda$ correlations.}
\end{table}

\newpage
\begin{table}[t]
\caption{Variational energies for $C_p$ = $W_0$ = 0.  All values are in
MeV.\,\,
Numbers in parentheses are statistical errors in the last digit.}
\begin{tabular}{lcddddc}
\hline   \\
 Clusters  && one-body & two-body     &  three-body      &  four-body	  &
Total\\
\\
\hline \\
Kinetic Energy$^{\rm a}$	&	&13.2(1)	&0.05(1)	&0.34(4)	&-0.09(2)	&13.5(1) \\
$\Lambda$N Potential	&$v_{0}(r)(1-\epsilon)$  &
															& 	-19.4(1)	&-0.72(4)		&0.03(5)		&-20.1(2) \\
	& $v_{0}(r)\epsilon P_{x}$	&&-8.2(1)	&0.56(2)	&0.18(2)	&-7.5(1) \\
&$\frac{1}{4} v_{\sigma} T^{2}_{\pi}  {\mbox{\boldmath$\sigma$}}_{\Lambda}
\cdot {\mbox{\boldmath$\sigma$}}_{N}$ &
&0.1	&-0.04	&0.01	&0.1 \\
$\Lambda$ Energy		&&13.2(1)	&-27.5(2)	&0.1(1)	&0.1(1)	&-14.0(1) \\
Nuclear Kinetic		&&82.9(4)	&49.5(2)	&8.0(2)	&-0.2(1)	&140.2(5) \\
$NN$ Potential	&$v_{ij}$  &&-147.5(4)	&-4.1(2)	&1.4(1)	&-150.1(4) \\
$NNN$ Potential &$V_{ijk}$  &&&-11.7(1)	&-3.2	&-12.0(1) \\
Nuclear Energy	&&82.9(4)	&-97.9(3)	&-7.8(1)	&0.8(1)	&-21.0(2) \\
Total Energy	&&96.1(4)	&-125.4(5)	&-7.7(1)	&0.9(1)	&-35.9(1) \\
\\
\hline
\end{tabular}
\tablenotetext[1]{Includes nucleon kinetic energy from $\Lambda$N
correlations.}
\end{table}

\end{document}